\documentclass[conference]{IEEEtran}
\IEEEoverridecommandlockouts

\usepackage{cite}
\usepackage{amsmath,amssymb,amsfonts}
\usepackage{algorithmic}
\usepackage{graphicx}
\usepackage{textcomp}
\usepackage{xcolor}
\usepackage{url}
\usepackage{comment}
\usepackage{enumitem}
\usepackage{xcolor}
\usepackage{colortbl}
\usepackage{cleveref}

\definecolor{mygreen}{rgb}{0,0.6,0}
\definecolor{mygray}{rgb}{0.5,0.5,0.5}
\definecolor{mymauve}{rgb}{0.58,0,0.82}
\definecolor{dkgreen}{rgb}{0,0.6,0}
\definecolor{gray}{rgb}{0.5,0.5,0.5}
\definecolor{mauve}{rgb}{0.58,0,0.82}
\definecolor{gray}{rgb}{0.4,0.4,0.4}
\definecolor{darkblue}{rgb}{0.0,0.0,0.6}
\definecolor{lightblue}{rgb}{0.0,0.0,0.9}
\definecolor{cyan}{rgb}{0.0,0.6,0.6}
\definecolor{darkred}{rgb}{0.6,0.0,0.0}
\definecolor{gray50}{gray}{.5}
\definecolor{gray40}{gray}{.6}
\definecolor{gray30}{gray}{.7}
\definecolor{gray20}{gray}{.8}
\definecolor{gray10}{gray}{.9}
\definecolor{gray05}{gray}{.95}

\newboolean{showcomments}
\setboolean{showcomments}{true} 
\ifthenelse{\boolean{showcomments}}
{\newcommand{\nb}[2]{
  \fcolorbox{black}{yellow}{\bfseries\sffamily\scriptsize#1}
  {\sf\small$\blacktriangleright$\textit{#2}$\blacktriangleleft$}
 }
 
}
{\newcommand{\nb}[2]{}
 
}

\newcommand{\resquestion}[2]{ %
	\vspace{5pt} %
	\noindent\fcolorbox{black}{blue!05}{%
		\parbox{0.97\linewidth}{%
			\textbf{RQ$_{#1}$.} \emph{#2} 
		}%
	}%
	\vspace{5pt} %
}%

\def\BibTeX{{\rm B\kern-.05em{\sc i\kern-.025em b}\kern-.08em
    T\kern-.1667em\lower.7ex\hbox{E}\kern-.125emX}}
    
\begin{document}

\title{Identifying and Replicating Code Patterns Driving Performance Regressions in Software Systems}

\author{\IEEEauthorblockN{Denivan Campos\IEEEauthorrefmark{1} Luana Martins\IEEEauthorrefmark{2}, Emanuela Guglielmi\IEEEauthorrefmark{1}, Michele Tucci\IEEEauthorrefmark{3}, \\ Daniele {Di Pompeo}\IEEEauthorrefmark{3}, Simone Scalabrino\IEEEauthorrefmark{1}, Vittorio Cortellessa\IEEEauthorrefmark{3}, Dario {Di Nucci}\IEEEauthorrefmark{2}, Rocco Oliveto \IEEEauthorrefmark{1}}

\IEEEauthorblockA{denivan.dasilva@unimol.it, lalmeida.martins@unisa.it, emanuela.guglielmi@unimol.it, michele.tucci@univaq.it,\\ daniele.dipompeo@univaq.it, simone.scalabrino@unimol.it, vittorio.cortellessa@univaq.it, ddinucci@unisa.it,\\rocco.oliveto@unimol.it}

\IEEEauthorblockA{
    \IEEEauthorrefmark{1}University of Molise, Pesche, Italy
    }
\IEEEauthorblockA{
    \IEEEauthorrefmark{2}University of Salerno, Salerno, Italy
    }
\IEEEauthorblockA{
    \IEEEauthorrefmark{3}University of L'Aquila, L'Aquila, Italy
    }
}

\maketitle

\begin{abstract}
Context: Performance regressions negatively impact execution time and memory usage of software systems. Nevertheless, there is a lack of systematic methods to evaluate the effectiveness of performance test suites. Performance mutation testing, which introduces intentional defects (mutants) to measure and enhance fault-detection capabilities, is promising but underexplored. A key challenge is understanding if generated mutants accurately reflect real-world performance issues.
Goal: This study evaluates and extends mutation operators for performance testing. Its objectives include (i) collecting existing performance mutation operators, (ii) introducing new operators from real-world code changes that impact performance, and (iii) evaluating these operators on real-world systems to see if they effectively degrade performance.
Method: To this aim, we will (i) review the literature to identify performance mutation operators, (ii) conduct a mining study to extract patterns of code changes linked to performance regressions, (iii) propose new mutation operators based on these patterns, and (iv) apply and evaluate the operators to assess their effectiveness in exposing performance degradations.
Expected Outcomes: We aim to provide an enriched set of mutation operators for performance testing, helping developers and researchers identify harmful coding practices and design better strategies to detect and prevent performance regressions.
\end{abstract} 

\begin{IEEEkeywords}
 Performance Issues, Performance Mutation Testing, Mining Software Repositories, Fault Injection.

\end{IEEEkeywords}

\section{Introduction}

Software performance issues are non-functional problems that reduce the efficiency, responsiveness, or scalability of software due to factors like slow processing, resource overuse, or system bottlenecks, ultimately harming user experience and wasting resources~\cite{olivo2015static, chen2020performance, delgado2021performance}.
These issues are challenging to detect and resolve because their occurrence often depends on variations in the execution environment, such as hardware configurations, workload patterns, or system settings.
Performance issues are usually detected by running performance tests (\textit{i.e.}, \textit{benchmarks}) with specific inputs to check whether performance metrics (\textit{e.g.}, execution time) degrade as the software evolves. However, the lack of reliable testing oracles, \textit{e.g.}, how slow should a computation be to be considered a performance issue, makes it challenging to identify and diagnose these issues consistently~\cite{nistor2013discovering,segura2017performance, sanchez2018performance}.

The literature highlights the critical role of performance testing in mitigating performance degradation and ensuring systems operate without issues~\cite{chen2017exploratory,chen2019inferring,traini2021software,Zhao2023_real_life_bugs,weng2023effective,imran2024empirical}. Despite its importance, there remains a significant gap in systematic methods for evaluating the effectiveness of performance test suites~\cite{TempleAJ21,imran2024empirical}. \looseness=-1

In this regard, performance mutation testing offers a promising approach. Like functional mutation testing, performance mutation consists of introducing intentional performance issues (\textit{mutants}) to assess the detection capabilities of performance tests.
S{\'a}nchez et al.~\cite{sanchez2018performance, sanchez2018search} and 
Delgado-P{\' e}rez et al.~\cite{delgado2021performance} explore the possible benefits and limitations of applying mutation testing for performance and define performance mutation testing (PMT). Like the classic mutation test, PMT aims to generate variants of the original program, \textit{i.e.}, performance mutants, where each variation simulates a performance error.
They also identified some challenges related to mutation testing and proposed seven new operators to model known patterns in C/C++ for inducing performance bugs. The authors compared the effectiveness of classical and performance mutants and found that the latter can be used to evaluate and improve the effectiveness of performance testing.

Despite its potential, PMT is still an under-explored field. A key challenge lies in the \emph{limited understanding of whether artificially generated mutants accurately reflect the performance issues in real-world systems}. More specifically, it is still unknown to what extent the generated mutants are relevant in different contexts. This gap limits developers' ability to evaluate the robustness of performance tests in identifying and mitigating actual performance regressions.

This registered report describes a study that evaluates and extends the mutation operators relevant to performance testing. The study begins with a comprehensive literature review to establish a catalog of existing performance mutation operators. The catalog will be enriched starting from the literature by introducing new operators derived analyzing real-world code changes that have demonstrably degraded system performance. To achieve this, we will leverage established benchmarks to identify such changes and extract patterns that will inform the design of novel mutation operators. Finally, we will evaluate the effectiveness of both existing and newly developed mutation operators on real-world systems, assessing their ability to simulate performance degradations. The evaluation will determine whether current performance tests can reliably detect the slowdowns introduced by these mutants, providing insights into their applicability across various contexts. \looseness=-1

Our findings are expected to advance performance mutation testing by providing an enriched set of mutation operators and empirical evidence of their impact on performance testing. The findings can assist in identifying harmful coding practices and design more effective strategies to detect and mitigate performance regressions, thus contributing to developing more robust and reliable performance testing methodologies.

The remainder of this paper is structured as follows. 
\Cref{section:RelatedWorks} presents the related work.
\Cref{section:Research_Design} presents the goal of the study and its research questions, which will be answered leveraging the methodology provided in \Cref{section:OurApproach}, and whose main threats to validity are reported in \Cref{section:ThreatsToValidity}.
Finally, \Cref{section:Conclusion} concludes the paper.

\section{Related Works}
\label{section:RelatedWorks}

This section describes the related work concerning mutation testing for bug replication and performance mutation testing.

\subsection{Mutation Testing for Bug Replication}
Just et al.~\cite{just2014mutants} investigated the correlation between detecting mutants generated using frequently used mutation operators and real defects. {\color{black}In particular, the authors used \textsc{Major} mutation framework provides the following set of mutation operators~\cite{Major2014}: \textit{Replace constants}, \textit{Replace operators}, \textit{Modify branch conditions}, and \textit{Delete statements}.} Furthermore, they examined whether code coverage could improve the efficacy of mutation analysis, showing a statistically significant correlation between mutant detection and actual failures, regardless of code coverage.

To determine whether classic mutation scores and fault detection efficacy are associated, Papadakis et al.~\cite{papadakis2018mutation} examined the relationship between mutation scores and real fault detection. They analyzed \textsc{C} and \textsc{Java} programs featuring real faults to demonstrate how the size of the test set affects the correlations between mutation scores and fault detection. {\color{black} To do so, the authors applied the mutation operators: \textit{Arithmetic, Logical, Conditional, and Relational Operators Replacement}, \textit{Operator Replacement Unary}, \textit{Statement Deletion}, and \textit{Literal Value Replacement}.} The results show that achieving higher mutation scores improves fault detection significantly and indicate that mutants provide good guidance for improving fault detection of test suites, although their correlation with fault detection is weak.

{\color{black}The existing performance mutation operators provide useful foundations for fault detection in various types of software testing.}
In the referred papers, Just et al.~\cite{just2014mutants} and Papadakis et al.~\cite{papadakis2018mutation} examine the relationship between traditional mutation analysis and actual fault detection, emphasizing the effectiveness of existing mutation operators and their correlation with fault detection in functional tests. {\color{black}While these operators are instrumental in verifying functional correctness, they do not directly address performance slowdowns related to suboptimal memory and resource management.}

Complementary, Wu et al.~\cite{wu2017memory} introduced memory mutation to replicate memory-related faults. {\color{black} They proposed memory mutation operators (e.g., \textit{Replace calloc with malloc, calloc with alloca, and malloc with alloca}, \textit{Remove null character assignment statement}, \textit{Replace dynamic memory allocation calls})} and compared them to traditional mutation operators. In addition, they addressed the problems associated with equivalent and duplicated mutants. Their results show that traditional operators insufficiently capture memory faults and reduce the effectiveness of test suites by 44\%.

{\color{black} While the memory mutation operators proposed by Wu et al.~\cite{wu2017memory} are valuable, they do not comprehensively simulate performance issues that involve memory management, data locality, thread synchronization, and cache usage.} In contrast, our study diverges by targeting performance tests, explicitly aiming to expand and refine mutation operators derived from real-world performance problems and evaluating their capability to produce realistic performance degradations.

\subsection{Performance Mutation Testing}
Delgado-P{\'e}rez et al.~\cite{delgado2021performance} investigated the application of performance mutation testing (PMT) to improve the detection of software performance bugs.
{\color{black}They proposed seven performance mutation operators related to execution time (e.g., \textit{Loop perturbation}, \textit{Method call}, \textit{Conditional statement}) and memory consumption (e.g., \textit{Object generation}, \textit{Collections})}. The authors then compared the effectiveness of performance mutation operations with traditional ones. The results show that performance mutants point out performance degradation while preserving the semantics of the original program to improve performance test effectiveness.

{\color{black}Jangali et al.~\cite{jangali2022automated} performed a comparison on the effectiveness of manually-written microbenchmarks with microbenchmarks generated with \textsc{JU2JMH} and other state-of-the-art tools in detecting real performance bugs.} In detail, it features five performance mutant operators {\color{black}(i.e., \textit{Primitive to Wrapper}, \textit{StringBuilder to StringBuffer}, \textit{StringBuilder to StringBuffer}, \textit{Swap of Operands in Condition}, and \textit{Simulation of Heavy-Weight Operations})} to evaluate the quality of microbenchmarks during performance bug detection. 
The results show that the \textsc{JU2JMH} benchmarks can cover more of the software applications than manually-written benchmarks.

Chen et al.~\cite{chen2022slowcoach} investigated whether synthetic bugs can be used to evaluate performance bug diagnosis tools, improve the quality of performance testing methodologies, and identify areas that need improvement in performance bug detection and localization approaches. They proposed a framework that leverages PMT to simulate software performance bugs and identify fault detection in C\textbackslash C++. {\color{black} Their performance operators (e.g., \textit{replace dfaisfast and fgrep calls with 0}, \textit{remove cache memoization}, \textit{remove early break from loops}, and \textit{prepend 1* loop to loop bodies})} were derived from real performance bugs and software optimizations, analyzing existing cases in the literature and real-world software projects.


{\color{black}These related studies~\cite{delgado2021performance, jangali2022automated, chen2022slowcoach} have introduced operators targeting specific aspects like execution time and memory consumption. However, they may not fully capture the wide variety of performance issues encountered in real-world applications. For example, a recent literature review~\cite{TANDEM2020} identifies three primary categories of performance bugs related to time, memory, and energy consumption. While the related works present operators for execution time and memory consumption, operators targeting energy consumption are still lacking. In addition, real-world performance bugs can arise from other factors besides data structures, such as redundant operations, misused algorithms, or inappropriate resource handling. Therefore, our study aims to collect existing performance mutation operators, propose new ones from the real world, and evaluate their effectiveness in identifying performance degradation.}

{\color{black}\subsection{Our paper}
Below, we highlight the key aspects that will set our study apart and enhance its insights.

\smallskip
\noindent
\textbf{Mining real-world performance issues.} Instead of relying on known causes of performance degradation from the literature, we will mine performance related issues from open-source projects. Specifically, we will manually analyze pairs of code before and after fixes to identify new performance mutation operators. We will work on performance issues confirmed as significant by developers from the respective projects, ensuring that the mutation operators we will define reflect real-world concerns rather than theoretical assumptions.

\smallskip
\noindent
\textbf{Validation through benchmarking.} We will run benchmarks on the commits before and after the fix to (i) further validate whether the problem was indeed a performance issue and (ii) confirm that the applied fixes improve performance. Our study brings a practical perspective on what developers consider important to fix as performance issues, offering insights on how to fix it based on real-world development practices.

\smallskip
\noindent
\textbf{Different research goals.} The referenced paper compares performance mutants against traditional mutants to evaluate their ability to induce perceptible performance degradation. In contrast, our goal is to identify and define new performance mutation operators that reflect real-world performance issues. We focus on discovering and systematizing these operators rather than comparing their impact against traditional mutants.

\smallskip
\noindent
\textbf{Different programming language.} The mutation operators discussed in related works, which simulate issues such as memory consumption, workload imbalances, or unnecessary operations, have primarily been applied to C/C++ programs. However, performance characteristics can vary significantly across programming languages. Therefore, we will consider these distinct characteristics and better simulate real-world performance issues in Java applications.

\smallskip
\noindent
\textbf{Broader Coverage of Performance Issues.} Unlike traditional functional testing~\cite{just2014mutants, papadakis2018mutation, wu2017memory}, performance testing requires different considerations and approaches. Previous works have proposed performance mutation operators focusing on execution time and memory consumption. We aim to propose new mutation operators that account for broader performance concerns (e.g., energy consumption), reflecting a more comprehensive view of real-world performance bugs.  
}

\section{Research Questions and Objectives}
\label{section:Research_Design}

The \textit{goal} of our empirical study is to define performance mutation operators based on real-world coding patterns and evaluate their effectiveness to inject realistic slowdowns across different contexts, {\color{black}i.e., specific code structure}. The \textit{purpose} of the study will be to (i) {\color{black} perform a literature review to identify a comprehensive set of performance mutation operators,} (ii) extract coding patterns associated with performance slowdowns {\color{black}in real-world project}, (iii) propose new performance mutant operators to stimulate the coding practices leading to performance slowdown, and (iv) identify whether the performance mutant operators can generate slowdowns in different contexts. By exploring the alignment between artificial mutants and genuine performance issues, this study will enhance our ability to evaluate and improve performance testing strategies and provide actionable insights for anticipating and mitigating performance regressions. Thus, the \textit{perspective} will be of both practitioners and researchers interested in understanding the coding patterns related to real-world performance issues to have practical insights and enhance their debugging and optimization workflows for detecting and resolving performance problems.
{\color{black}Our empirical investigation follows a sequential approach, addressing the following research questions (\textbf{RQ}):}

\resquestion{0}{Do developers use benchmarking to confirm performance issues and their resolution?}

The motivation behind \textbf{RQ$_0$} is twofold. First, we aim to examine how frequently developers rely on benchmarking to detect and validate performance issues. Second, when benchmarks are available, we will analyze whether their outcomes confirm the effectiveness of the applied fixes in resolving the identified issues. A lack of benchmarking for performance bugs or benchmarks that fail to demonstrate meaningful improvements may highlight critical gaps in current practices. {\color{black}In this case, we will generate the benchmarks to confirm the effectiveness of the applied fixes. Then, we will deepen our analysis by exploring the following research questions:} 

\resquestion{1}{What are the coding patterns related to performance issues in real-world software systems?}

\resquestion{2}{What performance mutation operators can simulate performance issues based on the identified coding practices?}

In \textbf{RQ$_1$}, we aim to uncover the coding patterns and root causes contributing to real-world software performance issues. 
We will construct a structured taxonomy that links specific coding practices to performance degradations by analyzing these patterns, providing a foundational understanding of how performance issues arise. Building on insights of \textbf{RQ$_1$}, with \textbf{RQ$_2$} we will focus on deriving performance mutation operators replicating the coding changes leading to performance issues. The identified set of mutation operators will enrich the existing catalog of performance mutants from literature. 

\resquestion{3}{To what extent are performance mutants valid substitutes for real-world performance issues in software performance testing?}

In \textbf{RQ$_3$}, we will validate the entire catalog of performance mutation operators—both those identified in the literature and the newly introduced ones. This evaluation will assess their generalizability across different contexts. {\color{black}Please, notice that we consider ``context'' as the specific code structure where a mutation operator is introduced. Since not all mutants can be applied in all contexts (e.g., a mutant that modifies a predicate clause requires a conditional expression), our methodology accounts for these constraints}. 
To refine this investigation, we pose two sub-questions:

\resquestion{3.1}{To what extent do performance mutants differ in induced slowdowns?}
\resquestion{3.2}{What is the influence of the injection context on the effectiveness of performance mutants?}

In \textbf{RQ$_{3.1}$}, we will evaluate the magnitude of slowdowns introduced by different mutation operators and their detectability within {\color{black} the same code structure (if applicable) to analyze their consistency and impact.}
~Meanwhile, \textbf{RQ$_{3.2}$} will extend this analysis to assess the consistency and variability of each mutation operator {\color{black} the same mutant operators across different code structures (if applicable) to determine its effectiveness and impact in varying contexts.}

We aim to advance the understanding of performance mutation testing by systematically addressing these research questions. We offer practical tools and insights to improve performance testing practices, anticipate performance issues, and validate comprehensive catalogs of performance mutation operators in real-world software systems.
\section{Methodology}
\label{section:OurApproach}

\begin{figure*}[th]
    \centering
\includegraphics[width=0.9\linewidth]{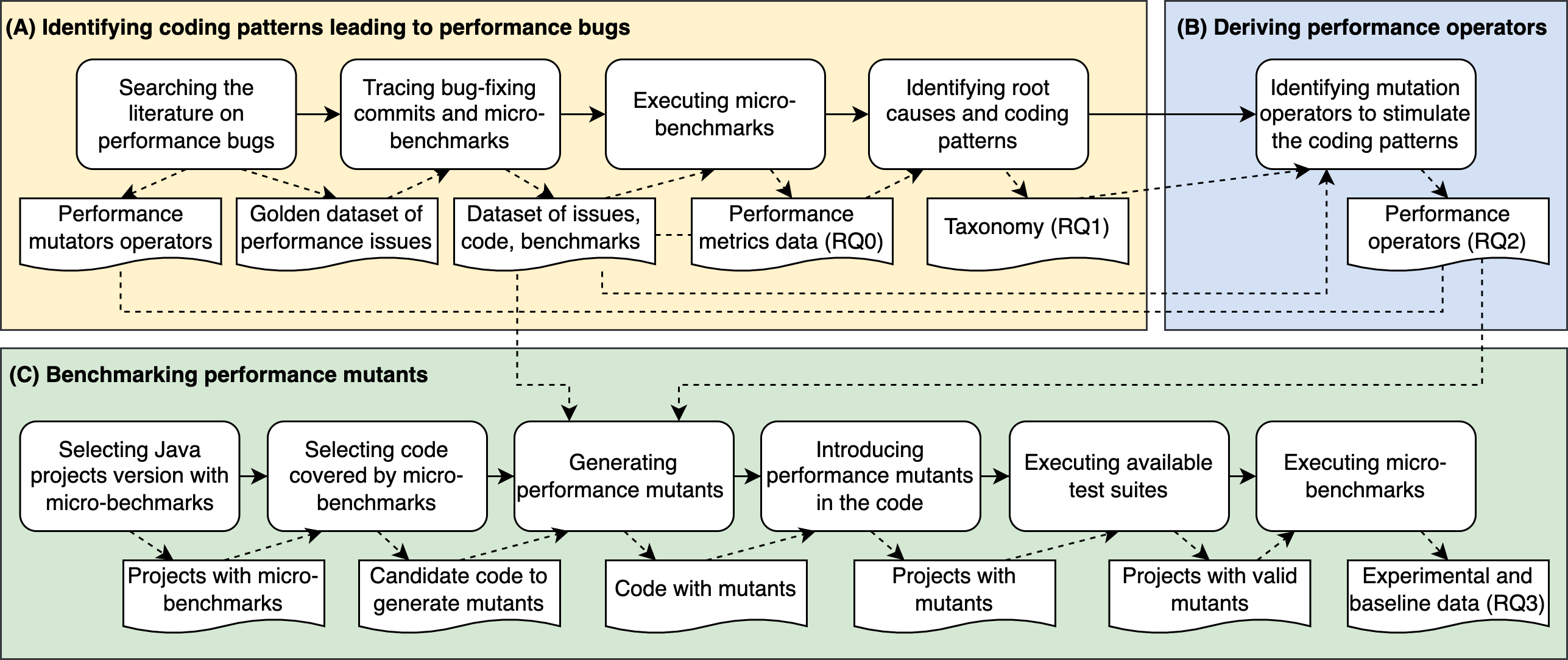}
    \caption{Overview of the Research Method.}
    \label{fig:designEMSE}
\end{figure*}

To design and report our empirical study, we adhere to the empirical software engineering guidelines outlined by Wohlin \textit{et al.}~\cite{Wohlin2012_Experimentation} and the ACM/SIGSOFT Empirical Standards\footnote{Available at: \url{https://github.com/acmsigsoft/EmpiricalStandards}}, particularly the standards for ``General Standard'', ``Repository Mining'', and ``Benchmarking''. \Cref{fig:designEMSE} provides an overview of our three-step study design, which we detail below.  

\subsection{Identifying Coding Patterns Leading to Performance Bugs}

\smallskip
\textbf{Searching the literature on performance issues.}
{\color{black} We conducted a preliminary literature review}
(discussed in \Cref{section:RelatedWorks}) to identify: (i) An initial set of performance mutation operators (\Cref{table:mutants_literature}); (ii) Curated datasets of performance-related issues.
We consolidated data from three manually curated datasets~\cite{Zhao2024_platform_agnostic, Zhao2023_real_life_bugs, Ding2020_pipeline_perf_tests}, resulting in {\color{black}497} performance-related issues extracted from Apache Software Foundation projects. This merged dataset forms our \textit{golden standard dataset}, which will serve as a starting point for analyzing and extending the catalog of performance mutation operators.

\begin{table*}[tb]
\footnotesize
\centering
\def \arraystretch{1.2}
\caption{Performance Mutation Operators from Literature}
\label{table:mutants_literature}
\begin{tabular}{p{1cm}p{4.5cm}p{10cm}p{0.6cm}}
\hline
\rowcolor{gray20}\textbf{Acronym} & \multicolumn{1}{c}{\textbf{Operator}} & \multicolumn{1}{c}{\textbf{Description}} & \textbf{Ref.}  \\ \hline

RCL & Removal of Stop Condition in Loop & RCL removes a stop condition in a loop to keep iterating until another condition is satisfied. &~\cite{delgado2021performance}  \\ 

\rowcolor{gray10}URV & Unnecessary Recalculation of Values & URV seeks for variables defined to store the value returned by an invocation to a method to force the recalculation of values. &~\cite{delgado2021performance}  \\ 

MSL & Move/Copy Statement into Loop & MSL searches for the generation of objects before a loop statement and moves it into the loop. &~\cite{delgado2021performance}  \\ 

\rowcolor{gray10}SOC & Swap of Operands in Condition & SOC swaps the operands in a condition linked by a binary logical operator (\&\& and $\parallel$) to evaluate the most time-consuming condition regardless of the other conditions. &~\cite{delgado2021performance}  \\ 

HWO & Simulation of Heavy-Weight Operation & HWO injects a delay right after each invocation to methods in third-party libraries and known heavy-weight operations (storage access, network connection). &~\cite{delgado2021performance} \\ 

\rowcolor{gray10}CSO & Creation of Short-lived Objects & CSO targets the methods that receive an object as a parameter and generates a clone of such objects, producing new short-lived objects every time the method is invoked. &~\cite{delgado2021performance}   \\ 

MSR & Memory Space Reservation & MSR modifies a collection with dynamic allocation to shrink or expand the reservation space for elements to simulate both cases. &~\cite{delgado2021performance}   \\ 

\rowcolor{gray10}PTW & Primitive to Wrapper & PTW replaces a primitive type (e.g., long) with its corresponding wrapper class (i.e., Long). &~\cite{jangali2022automated}  \\ 

STS & StringBuilder to StringBuffer & STS replaces a java.lang.StringBuilder object with a java.lang.StringBuffer. &~\cite{jangali2022automated} \\ 

\rowcolor{gray10}EFL & Enhanced For Loops & EFL replaces a traditional for-loop with a for-each loop to iterate over an array or a Collections class. &~\cite{jangali2022automated}  \\ \hline
\end{tabular}
\end{table*}

\smallskip
\textbf{Tracing bug-fixing commits and microbenchmarks.} 
The golden standard dataset includes bug descriptions but lacks corresponding code or benchmarks. To address this, we will trace each issue to its bug-fixing commit and associated benchmarks using \textsc{PyDriller}~\cite{Spadini2018}. We will identify benchmarks targeting the affected code by mining the modified files and methods in each commit. {\color{black} Microbenchmarking is a performance testing technique that evaluates the efficiency of small, isolated code snippets. It operates at a fine-grained level, similar to unit testing, and helps assess the impact of specific code changes on execution performance. Using microbenchmarking will help us confirm whether the code altered to fix a performance issue is actually to solve it~\cite{laaber2018evaluation}.}
Issues without identifiable bug-fixing commits or benchmarks will be excluded from our analysis.\looseness=-1

\smallskip
\textbf{Executing microbenchmarks.}
For issues with valid bug-fixing commits and benchmarks, we will execute the benchmarks on both the bug-fixing and predecessor commits to collect performance metrics {(i.e., \color{black}\textit{execution time}, and \textit{memory usage}}). Issues will be excluded if: (i) The benchmarks fail to build or run; (ii) The predecessor commit is unavailable or incompatible with the benchmarks. {\color{black} Please note that if the benchmarks to validate the performance issues in the \textit{Golden Dataset }are not available, we will create the benchmarks with the Ju2JMH, a tool that converts \textsc{JUnit} tests into \textsc{JMH} microbenchmarks~\cite{jangali2022automated}. This step will produce a dataset of performance metrics detailing the execution time of benchmarks before and after fixes, enabling analysis for \textbf{RQ$_0$}. As for \textbf{RQ$_3$}, we will extend the dataset of projects to evaluate the proposed mutation operators in real-world scenarios. Therefore, we will execute the benchmarks in bug-fixing and predecessor commits to collect performance metrics {(i.e., \color{black}\textit{execution time}, \textit{memory usage}, and \textit{Mutation Score}}.} 

\smallskip
\textbf{Identifying root causes and coding patterns.} The code pairs (before and after fixes) associated with performance issues in the dataset will be analyzed to identify root causes and coding patterns. {\color{black} We will use an LLM, \textsc{GPT-4o}, as an assistant to automate the process of identifying potential coding patterns associated with the performance issues. According to MacNiel et al~\cite{macneil2023experiences}, GPT-like LLMs can create explanations for the code, reducing the cognitive demand to understand complex codes from different systems. In addition, according to Colavito et al.~\cite{colavito2024leveraging}, \textsc{GPT}-like LLMs demonstrate a high level of agreement with human annotators, which could support us classifying the issues. 
Therefore, instead of asking GPT4-o to define recurring practices outright, we will have preliminary classifications and explanations, which we will manually validate across multiple instances. In particular, two experts will review and perform an open-coding of the explanation given by the GPT-4o regarding the potential root causes and coding patterns to answer \textbf{RQ$_1$}. }

\subsection{Deriving Performance Operators}


{\color{black} Using the taxonomy of coding patterns from \textbf{RQ$_1$}, we will employ \textsc{GPT-4o} to suggest mutation operators that simulate performance issues. The rationale behind using fix explanations to inform mutation operators is based on the intuition that fix patterns provide direct insight into recurring code transformations needed to resolve performance issues. It will use the coding patterns identified in the taxonomy to suggest potential code transformations that could degrade performance. For example, if the taxonomy identifies a coding pattern related to the use of collections, then \textsc{GPT-4o} could propose mutation operators that replaces data structures with a less efficient alternative, such as replacing HashMap with a LinkedHashMap. Then, similar to the previous step, two experts will analyze the GPT4-o answers to propose the final set of performance mutation operators.} The newly generated operators will be merged with the catalog of performance mutation operators identified in the literature (\Cref{table:mutants_literature}). This enriched catalog will answer \textbf{RQ$_2$}, offering a comprehensive set of operators for simulating real-world performance issues. 


\subsection{Benchmarking Performance Mutants}
{\color{black}
This section describes how the benchmarking performance mutants will be derived.
}

\smallskip
\textbf{Selecting \textsc{Java} projects with microbenchmarks.} 
~Mining all GitHub is unfeasible, given the number of projects it hosts. Therefore, we will rely on the \textsc{SEART} GitHub Search Engine\footnote{Available at \url{https://seart-ghs.si.usi.ch/}} to filter projects with at least 100 stars from the set of 106,018 \textsc{Java} projects in GitHub. We will also retain only projects where developers defined microbenchmarks for performance assessment. While other microbenchmarking tools are available (e.g., Caliper, Japex, or JUnitPerf), they are either less popular than the \textsc{Java Microbenchmarking Harness} (\textsc{JMH})\footnote{Available at \url{https://openjdk.java.net/projects/code-tools/jmh/.}}, discontinued, or not executable in an automated way~\cite{Leitner2017_AnExploratoryStudy,Stefan2017_UnitTestingPerformance}. Therefore, we will query the \textsc{Java} projects looking for those featuring (i) Maven as the dependency manager (i.e., contain a file named pom.xml) and (ii) a dependency on \textsc{JMH}, i.e., \texttt{<groupId>org.openjdk.jmh</groupId>}. As we do not aim to benchmark the performance evolution of the projects through their development, we will mine the last version of each project. In case of build failures, we will attempt to resolve them or select an older version, {\color{black} otherwise we will discard the project}. We will select one version because mining the entire history introduces overhead without adding value to synthetic data generation for method-level analysis~\cite{tip2024llmorpheus}. As a result, we will have a folder with the selected \textit{projects with microbenchmarks}.\looseness=-1

\smallskip
\textbf{Selecting code covered by microbenchmarks.}
We will use a lightweight instrumentation agent to identify the code executed during benchmarks to gather coverage data. Methods not covered by benchmarks will be excluded, ensuring our evaluation focuses on actively tested code.

\smallskip
\textbf{Generating performance mutants.} 
Using \textsc{GPT-4o}, we will generate mutants by applying controlled transformations to degrade performance. {\color{black}According to Wang et al.~\cite{wang2024exploratory}, mutations generated by \textsc{GPT-4o} present the highest precision in detecting real functional bugs in comparison to other closed-source LLMs from \textsc{GPT} family and popular open source LLMs as \textsc{DeepSeek-Coder-V2-236b} and \textsc{CodeLlama-Instruct-13b}.} Initially, we will follow the guidelines for \textit{prompt design} when generating mutations~\cite{wang2024exploratory, ma2024llmparser}. Our prompt will include:
(i) \textit{Instructions}, directing the LLM to generate mutants for the target code element;
(ii) \textit{Context}, clarifying that ``mutant'' refers to performance mutation and providing information such as the \textsc{Java} method (code snippets) and few-shot examples of real-world performance issues;
(iii) \textit{Input Data}, specifying the target code element and the number of mutants to generate; and
(iv) \textit{Output Indicator}, defining the \textsc{JSON} format for mutation outputs.
Mutants will be validated to ensure: (i) They compile successfully; (ii) They do not introduce functional bugs (validated through existing test suites). \looseness=-1

\smallskip
\textbf{Introducing performance mutants in the code.} Introducing performance mutants involves applying controlled code transformations that intentionally degrade performance, simulating real-world performance issues.
Some performance mutants LLMs generate might (i) not compile or (ii) introduce functional bugs. Therefore, we will \textbf{execute the available test suites} after the injection of every mutant.

\smallskip
\textbf{Executing microbenchmarks on the mutation.} 
We will execute benchmarks on both the original and mutated versions of the code. Performance slowdowns caused by the mutants will be recorded as \textit{experimental data} and compared to the \textit{baseline data} (original code) for analysis in \textbf{RQ$_3$}.

\subsection{Data Analysis}

In \textbf{RQ$_0$}, we aim to understand if developers use benchmarking to confirm performance issues. To this aim, we will present the percentage of issues with microbenchmarks that cover the code changed by the bug-fixing commit. {\color{black} The percentage will be calculated based on the performance issues covered by microbenchmarks.}
~Complementary, we will investigate the effectiveness of the fixes in solving the performance issue. We define the following null hypothesis:

\begin{description}
    \item[\textbf{H$_{0.0}$.}] There is \emph{no significant difference} in performance before and after the application of the fixes.
\end{description}

As for \textbf{RQ$_1$} and \textbf{RQ$_2$}, the validation procedure is the same. {\color{black}GPT-4o will provide preliminary classifications and explanations for root-causes, coding patterns and mutation operators. Then, two experts will review and perform an open-coding of the explanation given by the GPT-4o. Therefore, we will calculate the Kappa statistics~\cite{Cohen1960} to assess the experts' reliability. A Kappa coefficient of 0.8 or higher indicates a strong level of agreement between the two experts, demonstrating the reliability of the classification process. In addition, we will have discussions with them to solve disagreements.}

As for \textbf{RQ$_3$}, we first formulate the working hypotheses that we will later statistically assess. In \textbf{RQ$_{3.1}$}, we investigate the impact of different performance mutation operators within a specific context, {\color{black}i.e., the specific code structure where a mutation operator is introduced.} Differently, in \textbf{RQ$_{3.2}$}, we analyze the impact of a specific mutation operator across a set of contexts. Therefore, the null hypothesis for each RQ is the following:

\begin{description}
    \item[\textbf{H$_{0.1}$.}] There is \emph{no significant difference} in the impact of different mutation operators in a specific context.

    \item[\textbf{H$_{0.2}$.}] There is \emph{no significant difference} in the impact of a specific mutation operator across different contexts.
\end{description}

If one of the null hypotheses is statistically rejected, we will accept the corresponding alternative hypothesis, namely:

\begin{description}
    \item[\textbf{H$_{a.0}$}.] There is \emph{a significant difference} in performance before and after the application of the fixes.
    \item[\textbf{H$_{a.1}$.}] There is \emph{a significant difference} in the impact of different mutation operators in a specific context.
    \item[\textbf{H$_{a.2}$.}] There is \emph{a significant difference} in the impact of a specific mutation operator across different contexts.
\end{description}

We proposed to focus on execution time because microbenchmarking is usually targeted just at this metric (or related metrics like throughput). CPU usage and memory consumption could also be measured in microbenchmarking by employing profilers to attach to the JVM during the execution. However, this usually results in a noticeable overhead in execution time. Therefore, both measuring execution time and profiling microbenchmarks may alter the results. An alternative for RQ0 could be to focus on different performance metrics based on the reported issue (i.e., if an issue reports an increase in memory usage, we only profile the relevant microbenchmarks to measure memory usage, not execution time). As we are not executing the performance mutation testing in the RQ0, the Mutation Score metric does not apply.

\smallskip 
\textbf{Statistical modeling for RQ$_0$ and RQ$_3$.} We will produce paired distributions to answer both research questions. {\color{black} For $RQ_0$ we will focus on different performance metrics based on the reported issue (i.e., if an issue is reporting an increase in memory usage, we only profile the relevant microbenchmarks to measure memory usage, and not execution time). As we are not executing the performance mutation testing in the $RQ_0$, the \textit{Mutation Score} metric does not apply. Therefore, we will have different paired distributions for $RQ_0$ referring to the execution time (in seconds) and memory usage (bytes) calculated in the bug-fixing commit and its predecessor. For $RQ_3$ the paired distributions refer to the execution time (in seconds), memory usage (bytes), and \textit{Mutation Score} calculated in the baseline project and its mutated version. In the context of performance testing, the \textit{Mutation Score} quantifies the proportion of \textit{perf-mutants} that exhibit statistically significant slowdowns compared to the baseline, i.e., a \textit{perf-mutant} is killed if its execution time is statistically larger than the baseline.} In both RQs, we will use the non-parametric method based on the bootstrap approach~\cite{davison_bootstrap_1997} proposed by Kalibera and Jones~\cite{kalibera_quantifying_2020} to rigorously assess the statistical significance of differences between the paired distributions and the magnitude of the observed effects.

To determine whether the difference between the two distributions is statistically significant, we will construct a confidence interval for the ratio of means. The hierarchical bootstrap method generates multiple resampled distributions from the original data, maintaining the hierarchical structure of the experiment (e.g., repeated runs, multiple forks, execution variability). We compute the ratio of means for the paired distributions for each bootstrap iteration. If the $95\%$ confidence interval for the ratio of means includes the value 1, we conclude that the difference is not statistically significant. Conversely, if the confidence interval excludes 1, this provides evidence that the difference is statistically significant with $95\%$ confidence.
The magnitude of the effect will be computed as the point estimate of the ratio of means, which is calculated as the ratio of the arithmetic means of the two distributions. Along with the confidence interval, this provides a clear measure of the size of the difference and its associated uncertainty. For example, we might report that the metric in one distribution is ``$5.5\% \pm 2.5\%$ faster'' than in the other, with $95\%$ confidence.

\subsection{Publication of the Generated Dataset}
\label{subsection:Publication_dataset}
The golden dataset of performance issues, corresponding code pairs (before and after fixing the performance issue), and generated mutants will be publicly available online~\cite{Our_repo}. We also plan to release the data collection and analysis scripts that we will use to perform this study.

\section{Threats to validity}
\label{section:ThreatsToValidity}
This section discusses the potential threats that may affect the validity of our empirical study plan.

\smallskip
\textbf{Construct validity.} This validity has three main threats that we will attempt to mitigate. The first threat concerns the criteria we will use to select software projects. Despite the efforts to standardize the building process, we might still fall into build failures. We will attempt to diagnose the reasons for the failure and try to fix them manually. We will discard the project if we cannot fix the build failure. Another threat concerns the performance metric selection. We will benchmark the performance in terms of execution time before and after applying the mutations in the code. However, other performance metrics (e.g., CPU usage and memory consumption) could be useful in evaluating performance. 

\smallskip
\textbf{Internal validity.} 
Monitoring performance is challenging due to measurement noise~\cite{jangali2022automated}.
To mitigate this issue in RQ$_0$ and RQ$_3$, we will employ a state-of-the-art methodology~\cite{kalibera_quantifying_2020} to compare the paired distributions of execution times obtained for the bug-fixing commit and its predecessor. 
While we will manually validate the coding patterns and mutation operators for RQ$_1$ and RQ$_2$, their relevance and accuracy depend on experts' subjective judgment. We will measure their agreement level to assess the reliability of their validation process.

\smallskip
\textbf{External validity.} The main threat concerning the generalizability of the results is the selection of subject projects. {\color{black} We will select 
open-source \textsc{Java} projects,} which are only a fraction of the complete picture of open-source \textsc{Java} projects. Therefore, replications of this study on a larger number of projects and in different contexts would corroborate our findings. Therefore, we will provide all materials and scripts used in this study to stimulate further research~\cite{Our_repo}.

\smallskip
\textbf{Conclusion validity.} The main threat concerns the potential overlap between the evaluation subjects and the LLM prompt examples. Specifically, the LLM responses might not represent its ability to generate novel performance issues if the code used as input to generate the mutants were previously served as an example for the LLM.
We will mitigate this risk by including several projects besides the Apache Foundation projects composing the \textit{golden standard dataset} of performance issues. Additionally, we will ensure that the examples used in the prompt to guide the LLM do not overlap with the code to evaluate.
{\color{black} Another threat concerns the factors that can significantly influence the model's responses. While the context length might lead the model to struggle to maintain coherence, the generation of multiple mutations may lead the model to overfit on particular patterns. We will use well-defined prompt structures to mitigate these threats, limit the context to relevant information, and filter out invalid outputs.}

\section{Conclusions}
\label{section:Conclusion}

The first goal of our study is to understand whether developers rely on performance benchmarks to identify performance slowdowns and verify whether the fixings solve the issue. After this step, we will identify and validate performance mutation operators to support developers {\color{black}(i) evaluating the effectiveness of their performance tests, (ii) understanding whether a code change should be benchmarked, and (iii) developing static analysis tools to detect performance issues by detecting the patterns.} We will conduct this study on open source \textsc{Java} projects containing \textsc{JMH} microbenchmarks. We will start by collecting a golden standard dataset for proposing the mutation operators, and expand our analysis to validate them into a large dataset of performance issues. Then, we will employ statistical approaches to address the goals of our investigation and, finally, provide actionable implications for researchers and practitioners.
As an outcome of our exploratory study, we expect to provide the following key contribution:

\begin{itemize}
    \item a taxonomy of causes and coding practices leading to performance issues and a catalog of new performance mutation operators to identify the issues;
    \item evidence of the usefulness of performance tests and the impact of performance mutation operators to improve performance testing in different software contexts;
    \item a novel and curated dataset of performance issues and their corresponding pairs of code pre- and post-fix;
    \item an online appendix that will provide all material and scripts employed to address the goals of the study.
\end{itemize}


\bibliographystyle{IEEEtran}
\bibliography{references}

\end{document}